\documentclass[pra,twocolumn,showpacs,preprintnumbers,amsmath,amssymb]{revtex4}
\usepackage[english]{babel}
\usepackage{indentfirst}
\usepackage{epsfig}
\usepackage{amsfonts,amssymb,latexsym,amsmath,enumerate}
\usepackage{bm}

\def\diag{{\rm diag}}
\def\Tr{{\rm Tr}}

\begin{document}


\title{Radiative polarization of electrons in a strong laser wave}
\author{Dmitry V. Karlovets}
\email{d.karlovets@gmail.com}
\affiliation{Tomsk Polytechnic University, Lenina 30, 634050, Tomsk, Russian Federation}

\date{\today}

\begin{abstract}
We reanalyze the problem of radiative polarization of electrons brought into collision with a circularly polarized strong plane wave. We present an independent analytical verification of formulae for the cross section given by D.\,Yu. Ivanov et al [Eur.\ Phys.\ J. C \textbf{36}, 127 (2004)]. By choosing the exact electron's helicity as the spin quantum number we show that the self-polarization effect exists only for the moderately relativistic electrons with energy $\gamma = E/mc^2 \lesssim 10$ and only for a non-head-on collision geometry. In these conditions polarization degree may achieve the values up to $65\,\%$, but the effective polarization time is found to be larger than 1\,s even for a high power optical or infrared laser with intensity parameter $\xi = |{\bf E}| m c^2/E_c \hbar \omega \sim 0.1$ ($E_c = m^2 c^3/e \hbar$). This makes such a polarization practically unrealizable. We also compare these results with the ones of some papers where the high degree of polarization was predicted for ultrarelativistic case. We argue that this apparent contradiction arises due to the different choice of the spin quantum numbers. In particular, the quantum numbers which provide the high polarization degree represent neither helicity nor transverse polarization, that makes the use of them inconvenient in practice.
\end{abstract}




\maketitle

\section{\label{1}Introduction}

The process of emission of a photon by an electron moving in the field of a plane wave of arbitrary strength was considered in quantum electrodynamics (QED) at the beginning of 1960's \cite{Nikishov-1_JETP_64, G, Nikishov-3_JETP_64}. Polarization effects in this process were studied in the limiting case of a weak wave (ordinary linear Compton scattering) e.g. in Refs. \cite{BLP, L-T, T, Serbo_NIMA_98} and in the general non-linear regime (i.e. at the tree-level in a Furry picture) e.g. in Refs. \cite{Ternov, G-2, Gr, Pol}. The complete description of the polarization effects in the non-linear case was given more recently in Ref.\cite{Serbo_EPJ_04}. Polarization states of the initial and final particles are of importance for a number of processes studied at the current accelerators (such as LHC; see e.g. \cite{Piazza, Muller}) and planned for studying at the future colliders (such as CLIC and ILC; see e.g. \cite{Artru, Zimm, B}). 

The possibility to use the linear and non-linear Compton scattering for production of the polarized electron (positron) beams was discussed in Refs.\cite{Klimenko, Bagrov_Preprint, B1, B2, Ternov-90, Serbo_PR_03, Pap} and some others. The general conclusion on this topic was made in Ref.\cite{Serbo_PR_03} where the absence of the longitudinal radiative polarization during a head-on collision of the ultrarelativistic electrons and the optical photons was demonstrated. In fact, such a conclusion was made for the first time as early as 1968 by the Ternov group in Ref.\cite{Ternov} soon after discovery of the self-polarization effect in synchrotron radiation. Nevertheless, later it was pointed out by the same group that there exist some spin operators allowing one to find the directions that are preferable for the electron (positron) spin, so the radiative polarization may take place \cite{Bagrov_Preprint, B1, B2}. Therefore, the problem cannot be considered to be finally solved.

In the present paper we analytically calculate the cross section of this process for arbitrary polarized initial and final electrons, with a circularly polarized laser wave and unpolarized final photon. The laser wave is assumed to be circularly polarized (rather than linearly polarized) because it provides the highest influence on the electron's spin (see e.g. $\S $87 in \cite{BLP}). The results of our calculations completely coincide with those obtained via computer code in Ref.\cite{Serbo_EPJ_04}, but the final expression for the squared amplitude is presented in the form which allows one to study the case of a moderately relativistic electron and a non-head-on collision.

It is important to note that the zero result of Refs.\cite{Ternov, Serbo_PR_03} was obtained when neglecting the terms of the order of $\gamma^{-1} \ll 1$ and $\theta_e \ll 1$ in the cross section ($\theta_e$ is the electron's scattering angle); see the corresponding footnote in \cite{Serbo_PR_03}. In the present paper we systematically take into account all the small terms in the amplitude and estimate their influence on the degree of polarization. The last becomes possible because we use the exact electron's helicity as the spin quantum number instead of the Lorentz-invariant quantity $\zeta^{\prime}_3$ (used e.g. in \cite{Serbo_PR_03}) which coincides with the final electron's double mean helicity with an accuracy $\mathcal O(\gamma^{-2})$ (see in more detail \cite{Serbo_EPJ_04}). In particular, we will show that the total probabilities of the process with spin flips, $W_{\uparrow \downarrow }$ and $W_{\downarrow \uparrow }$, are equal to each other exactly with that accuracy. Therefore, the polarization degree is negligibly small already at $\gamma \sim 10^2$. However, if the particles collide non-head-on and the electron's energy is not too high, the degree of polarization may turn out to be noticeably higher. We study in detail the case of a non-head-on collision of the moderately relativistic electrons with the laser photons and show that degree of the longitudinal polarization may reach the values of $P \lesssim 65\,\%$ for electrons with $\gamma \lesssim 10$ and the small angles of collision: $\alpha \ll \pi/2$ ($\alpha = \pi$ for head-on geometry).

Besides that, we study the influence of the laser wave strength on the polarization degree and find no significant deviations from the linear Compton scattering case for parameters of the laser similar to those of the well-known experiment conducted at SLAC \cite{SLAC}. On the other hand, it turns out that only the high power lasers provide a reasonable polarization time (which is $\sim 1$\,s for an optical laser with $|{\bf E}|/E_c \gtrsim 10^{-6}$), since the probability of the process with a spin flip is extremely low in the linear regime for small angles of collision. This circumstance makes the experimental realization of such a polarization technique practically impossible.

Finally, we explain why the results of Ref.\cite{Serbo_PR_03} and Refs.\cite{Bagrov_Preprint, B1, B2} do not agree. In brief, the different spin quantum numbers were used in these papers.
In particular, if one chooses the electron's helicity as the spin quantum number the radiative polarization would be absent in the relativistic case (in accordance with \cite{Serbo_PR_03}).

The paper is organized as follows. After general definitions given in Sec.\ref{sec1} we proceed to the calculation of the squared amplitude summed over polarizations of the final photon in Sec.\ref{Sect2}. The general expression for the squared amplitude, its comparison with results of Ref.\cite{Serbo_EPJ_04} and the final electron's polarization are presented in Sec.\ref{Sect3}. Analysis of the process with a spin flip both in the linear and non-linear regimes is performed in Sec.\ref{Sect4}. Our conclusions are presented in Sec.\ref{Sect5}. Appendix A contains some details of the squared matrix element calculations. Appendix B contains an expression for the squared matrix element written in the limiting case of a weak wave (linear Compton scattering) and its comparison with results of Refs.\cite{L-T, T}. A system of units $\hbar = c = 1$ is used throughout the text. The scalar product of 4-vectors is defined with the use of the metric $g^{\mu \nu} = \diag(1,-1,-1,-1)$, so that $kr \equiv k^{\mu}r_{\mu} = \omega t - ({\bf k} {\bf r})$.

\section{\label{sec1}Kinematics}

\subsection{\label{sec1.1}General definitions and conservation laws}

We consider the process of emission of a photon with momentum $k^{\prime}$ by an electron with kinetic momentum $p$ moving in the field of a plane wave possessing arbitrary strength:
\begin{eqnarray}
\displaystyle \gamma(n k) + e^-(q) \rightarrow e^-(q^\prime) + \gamma(k^\prime), \label{Eq1}
\end{eqnarray}
with 
\begin{eqnarray}
\displaystyle q^{\nu} = p^{\nu} - k^{\nu} \frac{e^2 A^2}{2 (pk)},\ {q^{\nu}}^{\prime} = {p^{\nu}}^{\prime} - k^{\nu} \frac{e^2 A^2}{2 (p^{\prime}k)}\label{Eq1a}
\end{eqnarray}
being the averaged quasi-momenta of electrons moving in the laser wave with 4-potential $A^{\mu} \equiv A^{\mu} (kr)$, and $n$ is an integer (a number of harmonic). 

According to the reaction (\ref{Eq1}), the emission of a photon with frequency $\omega^{\prime} = k^{\prime}_0$ occurs after absorbing $n$ photons with frequency $\omega = k_0$ from the wave. The potential of the wave possessing 100\,\% circular right-handed polarization can be represented as follows:
\begin{eqnarray}
\displaystyle A = \frac{m \xi}{e} \Big (e_1 \cos {(kr)} + e_2 \sin {(kr)} \Big ), \label{Eq2}
\end{eqnarray}
where $e_1, e_2$ are the space-like unit 4-vectors with the following properties: $e_i e_j = - \delta_{ij}, (e_i k) = (e_i k^{\prime}) = 0; i = 1,2$. These vectors can be constructed by using the momenta of the particles which enter the reaction (\ref{Eq1}) (see e.g. \cite{BLP}):
\begin{eqnarray}
& \displaystyle e_1^{\mu} = \frac{N^{\mu}}{\sqrt{-N^2}}, \ e_2^{\mu} = \frac{P^{\mu}}{\sqrt{-P^2}},\cr & \displaystyle P^{\mu} = q^{\mu} + {q^{\mu}}^{\prime} - (n k^{\mu} + {k^{\mu}}^{\prime}) \frac{(q^{\nu} + {q^{\nu}}^{\prime})(n k_{\nu} + {k_{\nu}}^{\prime})}{(n k + k^{\prime})^2}, \cr & \displaystyle N^{\mu} = \varepsilon^{\mu \nu \eta \rho}P_{\nu}(q_{\eta} - {q_{\eta}}^{\prime})(n k_{\rho} + {k_{\rho}}^{\prime}) = \cr & \displaystyle = 4n \varepsilon^{\mu \nu \eta \rho}p_{\nu}{k_{\eta}}^{\prime}k_{\rho} = 4n \varepsilon^{\mu \nu \eta \rho}{p_{\nu}}^{\prime}{k_{\eta}}^{\prime}k_{\rho}. \label{Eq2.5}
\end{eqnarray}
The last two identities are fulfilled due to the conservation laws. According to these definitions, the vector $e_1$ is orthogonal to all the momenta: $(e_1 p) = (e_1 p^{\prime})  = 0$, and the following equalities are fulfilled for the vector $e_2$:
\begin{eqnarray}
& \displaystyle (p e_2) = (p^{\prime} e_2) = - \sqrt{\frac{2 n (p^{\prime}k)(pk)}{(k^{\prime}k)} - m_\star^2}, \cr & \displaystyle \sqrt{-P^2} = - 2 (p e_2), \sqrt{-N^2} = - 4 n (k^{\prime}k)(p e_2). \label{Eq2.65}
\end{eqnarray}
Here, $m_\star^2 = m^2 (1 + \xi^2 ) \equiv q^2$ is the squared effective electron mass in the laser field. Therefore, the vectors $e_1, e_2$ (or rather their space components ${\bf e}_i$) describe the photon polarization perpendicular to the scattering plane and in that plane, respectively. Note that all calculations need only be performed for the right-hand polarization of the laser wave ($\xi_2 = + 1$). The process amplitude for the left-hand polarization ($\xi_2 = - 1$) can be obtained by changing the signs of the pseudo-scalar terms $M_s, M_{s^{\prime}}$ (see Sec.\ref{Sect2}).

The expression (\ref{Eq2}) also contains a ``classical'' dimensionless parameter of the laser wave strength (do not confuse with Stokes parameters):
\begin{eqnarray}
\displaystyle \xi = \frac{e \sqrt{- A^2}}{m} = \frac{e |{\bf E}|}{m\omega} = \frac{|{\bf E}|}{E_c}\frac{m}{\omega} \equiv \frac{|{\bf E}|}{E_c}\frac{m c^2}{\hbar \omega}, \label{Eq2.1}
\end{eqnarray}
where $|{\bf E}|$ is the root-mean-square electric field strength of the wave in the laboratory frame of reference, $E_c = m^2 c^3/e \hbar \simeq 1.3 \times 10^{16}$\,V/cm is the critical field of QED (the corresponding intensity is $I_c = E_c^2 c/4\pi \approx 0.5 \times 10^{30}\,W{cm}^{-2}$). The limiting case $\xi \ll 1$ corresponds to the weak field of the wave. In order to make an accurate transition to the linear Compton scattering case it is also necessary to change a normalization in (\ref{Eq2}) to the one photon in the unit volume. For this purpose one needs to put $\xi^2 = 4 \pi e^2/m^2 \omega$ and then define the cross section of the process. The alternative variant consists in the transition in the general non-linear case to the ``effective'' cross section staying finite in the limiting case $\xi \rightarrow 0$ (see below).

The energy-momentum conservation law for the reaction (\ref{Eq1}) may be written as
\begin{eqnarray}
\displaystyle n k + p + k \frac{m^2 \xi^2}{2 (pk)} = p^{\prime} + k^{\prime} + k \frac{m^2 \xi^2}{2 (p^{\prime}k)}. \label{Eq3}
\end{eqnarray}
Multiplying this equality by the vectors $p, p^{\prime}, k, k^{\prime}$, we obtain the following useful identities:
\begin{eqnarray}
\displaystyle & (pk) = (p^{\prime}k) + (k^{\prime}k), \cr & \displaystyle (p^{\prime}p) - m^2 =  (k^{\prime}k) \Big (n - \frac{m^2 \xi^2 (k^{\prime}k)}{2 (pk) (p^{\prime}k)}\Big ),\cr & \displaystyle (pk^{\prime}) = (p^{\prime}k) \Big (n - \frac{m^2 \xi^2 (k^{\prime}k)}{2 (pk) (p^{\prime}k)}\Big ), \cr & \displaystyle (p^{\prime}k^{\prime}) = (pk) \Big (n - \frac{m^2 \xi^2 (k^{\prime}k)}{2 (pk) (p^{\prime}k)}\Big ). \label{Eq4}
\end{eqnarray}
Taking into account the first equality, one can rewrite the conservation law (\ref{Eq3}) as
\begin{eqnarray}
& \displaystyle p + k  \Big (n - \frac{m^2 \xi^2 (k^{\prime}k)}{2 (pk) (p^{\prime}k)}\Big ) = p^{\prime} + k^{\prime}. \label{Eq5}
\end{eqnarray}

The vectors of the electron polarization $s, s^{\prime}$ are space-like, $(sp) = (s^{\prime}p^{\prime}) = 0$, and in laboratory frame of reference they are
\begin{eqnarray}
& \displaystyle s = \Big \{ \frac{({\bm \zeta} {\bf p})}{m}, {\bm \zeta} + {\bf p} \frac{({\bm \zeta} {\bf p})}{m (E + m)}\Big \} = \Big \{ \frac{|{\bf p}|}{m} \zeta_{\parallel }, \frac{{\bf p}}{|{\bf p}|}\gamma \zeta_{\parallel } + {\bm \zeta}_{\perp }\Big \},\cr
& \displaystyle s^{\prime} = \Big \{ \frac{({{\bm \zeta}}^{\prime} {{\bf p}}^{\prime})}{m}, {\bm \zeta}^{\prime} + {{\bf p}}^{\prime} \frac{({\bm \zeta}^{\prime} {\bf p}^{\prime})}{m (E^{\prime} + m)}\Big \} = \cr & \displaystyle = \Big \{ \frac{|{\bf p}^{\prime}|}{m} \zeta^{\prime}_{\parallel }, \frac{{\bf p}^{\prime}}{|{\bf p}^{\prime}|}\gamma^{\prime} \zeta^{\prime}_{\parallel } + {\bm \zeta}^{\prime}_{\perp }\Big \}. \label{Eq6}
\end{eqnarray}
Here, $E = \gamma m, \ {\bf p}, \ E^{\prime} = \gamma^{\prime} m, \ {\bf p}^{\prime}$ are the energies and momenta of the initial and final electrons, $\zeta_{\parallel } = 2 \lambda
_e, \zeta^{\prime}_{\parallel } = 2 \lambda^{\prime}_e$ are the double mean helicities of the electrons. Multiplying equality (\ref{Eq5}) by the 4-vectors of spins, we arrive at the ``spin conservation laws'':
\begin{eqnarray}
& \displaystyle (sk) \Big (n - \frac{m^2 \xi^2 (k^{\prime}k)}{2 (pk) (p^{\prime}k)}\Big ) = (sp^{\prime}) + (sk^{\prime}),\cr & \displaystyle (s^{\prime}k) \Big (n - \frac{m^2 \xi^2 (k^{\prime}k)}{2 (pk) (p^{\prime}k)}\Big ) = (s^{\prime}k^{\prime}) - (s^{\prime}p)\label{Eq7}
\end{eqnarray}
that will be used below.

\subsection{\label{sec1.2}Probability of the process and cross section}

The probability of the reaction (\ref{Eq1}) per second is defined as a probability of ``decay'' of the plane-wave electron with quasi-energy $q_0$ into two plane-wave particles (electron and photon) in the element of a phase volume $d^3 q^{\prime}d^3 k^{\prime}/(2\pi )^6$:
\begin{eqnarray}
& \displaystyle dW = \sum_{n=1}^{\infty}dW_n, \ dW_n = (2 \pi)^4 \delta^{(4)} (nk + q - q^{\prime} - k^{\prime})\cr & \displaystyle \times |M_n|^2 \frac{1}{2q_0 2q_0^{\prime}2\omega^{\prime}}\frac{d^3 q^{\prime}}{(2\pi )^3}\frac{d^3 k^{\prime}}{(2\pi )^3}. \label{Eq3.69}
\end{eqnarray}
It is useful to introduce the notion of the effective cross section by dividing the probability by the flux density of colliding particles $j$ (see in more detail \cite{Serbo_EPJ_04}):
\begin{eqnarray}
&& \displaystyle d\sigma = \sum_{n=1}^{\infty}d\sigma_n,\ d\sigma_n = \frac{dW_n}{j}, \ j = \frac{m^2 \xi^2}{4 \pi q_0 e^2}(pk) . \label{Eq3.70}
\end{eqnarray}
After elimination of the delta-function in (\ref{Eq3.69}) in the center-of-mass system ($n{\bf k} + {\bf q} = 0$), the formula for effective cross section gains the simple form:
\begin{eqnarray}
& \displaystyle d\sigma_n = \frac{r_e^2}{x}|\mathcal M_n|^2 dy d\varphi, \ |\mathcal M_n|^2 = \frac{1}{4 \pi e^2 m^2 \xi^2}|M_n|^2,\cr & \displaystyle y = \frac{(kk^{\prime})}{(pk)},\ x = \frac{2 (pk)}{m^2}\equiv \frac{2 (pk)}{m^2 c^2}.\label{Eq3.71}
\end{eqnarray}
Here, $\varphi$ is the common azimuthal angle of the scattering plane, and $r_e = e^2/m$ is the classical electron radius. For the following convenience, we have separated the factor $4 \pi e^2 m^2 \xi^2$ from the square of the matrix element. This representation, in particular, makes the comparison between the formulas derived and the ones for the linear scattering more convenient, since the squared matrix element $|\mathcal M_n|^2$ stays finite in the limiting case $\xi \rightarrow 0$.

When considering the process with a spin flip ($2 \lambda_e = \pm 1, 2\lambda^{\prime}_e = \mp 1$) in the general non-linear regime, it is necessary to use the formula for total (but not differential) probability per second:
\begin{eqnarray}
& \displaystyle W = \sum_{n=1}^{\infty}W_n, \ W_n = \frac{1}{T_0} \frac{\xi^2}{8 \pi} \frac{m}{q_0} \frac{x}{r_e^2} \sigma_n, \label{Eq3.73}
\end{eqnarray}
where $T_0 = 1/e^2 m \equiv \hbar^2/e^2 m c \simeq 1.765 \times 10^{-19}$ s. As the cross section $\sigma_n$ in (\ref{Eq3.71}) is invariant under the action of the Lorentz boosts along the collider axis, one can calculate it in an arbitrary frame of reference and then substitute into Eq.(\ref{Eq3.73}). 

Study of collisions between photons and electrons (or positrons) usually implies the head-on geometry and, accordingly, the use of formulas similar to Eq.(\ref{Eq3.71}) (see e.g. \cite{BLP, Serbo_NIMA_98, Serbo_EPJ_04, Heinzl_10}). The radiative polarization effect in this case is negligibly small (see \cite{Ternov, Serbo_PR_03} and Sec.\ref{Sect4} below). Therefore, our main focus will be on the non-head-on geometry and the moderately relativistic energies of electrons. In Sec.\ref{Sect4} we will use the laboratory frame of reference in which the electron performs a non-head-on collision with a laser photon at the angle $\alpha$. The final photon frequency in this frame is
\begin{eqnarray}
& \displaystyle \omega^{\prime} = \frac{n (pk)}{q^0 + n \omega - ({\bf n}^{\prime},{\bf q} + n{\bf k})}, \label{Eq3.73b}
\end{eqnarray}
where ${\bf n}^{\prime} = {\bf k}^{\prime}/\omega^{\prime}$ is denoted. The probability of the process per second in this frame is obtained from (\ref{Eq3.69}) as:
\begin{eqnarray}
& \displaystyle dW_n = \frac{1}{T_0} \frac{\xi^2}{4 \pi} \frac{m}{q_0 n x} \left ( \frac{\omega^{\prime}}{m}\right )^2|\mathcal M_n|^2 d\Omega, \label{Eq3.73a}
\end{eqnarray}
where the integration is performed over the final photon angles (see Eq.(\ref{Eq5.2a})). Note that the azimuthal angle $\varphi$ in (\ref{Eq3.73a}) does not coincide with the one in (\ref{Eq3.71}), since in the frame of reference being used the particles collide non-head-on. This implies an additional azimuthal dependence in (\ref{Eq3.73a}) which is absent for the head-on geometry and the circularly polarized laser wave.

If the probability of the process with a spin flip differs from zero, the time of the radiative polarization (relaxation time) is defined as follows:
\begin{eqnarray}
& \displaystyle T_{pol} = \frac{1}{W_{\uparrow \downarrow } + W_{\downarrow \uparrow }},\ W_{\uparrow \downarrow (\downarrow \uparrow )} \equiv W(2 \lambda_e = \pm1, 2 \lambda_e^{\prime} = \mp 1), \cr & \displaystyle W \equiv \sum \limits_{n=1}^{\infty} W_n. \label{Eq3.76}
\end{eqnarray}
According to the usual statistical interpretation, the portions of the final electrons in a beam, $n_{\uparrow (\downarrow )}$, being in the polarization states with $2 \lambda_e^{\prime} = \pm 1$ and the process asymmetry $P$ (degree of polarization) are found as:
\begin{eqnarray}
& \displaystyle n_{\uparrow (\downarrow )} = \frac{W_{\downarrow \uparrow (\uparrow \downarrow )}}{W_{\uparrow \downarrow } + W_{\downarrow \uparrow }},\ P = n_{\uparrow} - n_{\downarrow} = \frac{W_{\downarrow \uparrow } - W_{\uparrow \downarrow }}{W_{\uparrow \downarrow } + W_{\downarrow \uparrow }}. \label{Eq5.8b}
\end{eqnarray}
The complete polarization of the electron beam with a degree $P$ comes when the time interval $\Delta t$ of its flight through a laser flash is $\Delta t \gg T_{pol}$. 

Finally, note that all these definitions refer to the emission rate calculated for only one electron that implies neglecting the collective effects in the electron beam (\textit{incoherent} regime of emission; see e.g.\,Eq.(74) in \cite{Heinzl_10}). Actually, even in the well-studied problem of the self-polarization process due to the synchrotron radiation the rigorous theory of the collective spin-effects is absent. We will return to these problems in Sec.\ref{Sect5}.

\section{\label{Sect2}The square of the process matrix element}

At the tree-level in the Furry picture the squared matrix element of a photon emission by a plane-wave electron moving in the quasi-classical field of the laser wave with potential (\ref{Eq2}) has the following form (notations of \cite{BLP} are used):
\begin{eqnarray}
\displaystyle |M_n|^2 = 4\pi e^2  \Tr \{\rho_{(e)}^{\prime}Q^{\mu}\rho_{(e)} {\bar Q}^{\nu} {\rho^{(\gamma)}}^{\prime}_{\nu \mu}\}. \label{Eq3.78}
\end{eqnarray}
Here, $\rho_{(e)}, \rho_{(e)}^{\prime}, {\rho^{(\gamma)}}^{\prime}_{\nu \mu}$ are the density matrices of the initial and final electrons and the final photon, respectively. Besides, in (\ref{Eq3.78}) the following notations are used:
\begin{eqnarray}
& \displaystyle Q = \sum \limits_{j=1}^{3} g_j G_j, \ g_1 = J_n,\ g_2 = \frac{i}{2}(J_{n+1} - J_{n-1}),\cr & \displaystyle g_3 = \frac{1}{2}(J_{n+1} + J_{n-1}),\ G_1 = \gamma + k (\gamma k) \frac{m^2 \xi^2}{2 (p^{\prime}k)(pk)},\cr & \displaystyle G_2 = \gamma (\gamma k) (\gamma e_1) \frac{m \xi (k^{\prime}k)}{2(p^{\prime}k)(pk)} + \frac{m\xi}{(pk)} (k (\gamma e_1) - e_1(\gamma k)), \cr & \displaystyle G_3 = \gamma (\gamma k) (\gamma e_2) \frac{m \xi (k^{\prime}k)}{2(p^{\prime}k)(pk)} + \cr & \displaystyle + \frac{m\xi}{(pk)} (k (\gamma e_2) - e_2(\gamma k)), \label{Eq3.79}
\end{eqnarray}
where $\gamma \equiv \gamma^{\mu}$ are the Dirac matrices, $\bar Q = \gamma^0 Q^{\dagger } \gamma^0$, and $J_n \equiv J_n(z_n)$ is the Bessel function depending on the invariant argument
\begin{eqnarray}
\displaystyle z_n = \frac{m \xi (k^{\prime}k)}{(p^{\prime}k)(pk)} \sqrt{\frac{2 n (p^{\prime}k)(pk)}{(k^{\prime}k)} - m_\star^2}. \label{Eq3.3}
\end{eqnarray}
We will not study the polarization of the final photon, so one may put ${\rho^{(\gamma)}}^{\prime}_{\nu \mu} = -g_{\nu \mu}/2 $ or, equivalently, ${\rho^{(\gamma)}}^{\prime}_{\nu \mu} = (e_{1 \nu} e_{1 \mu} + e_{2 \nu} e_{2 \mu})/2$ (with vectors $e_1, e_2$ defined in (\ref{Eq2.5})) and then double the squared amplitude.

The squared scattering amplitude (\ref{Eq3.78}) (as well as $|\mathcal M_n|^2$) may be represented as follows:
\begin{eqnarray}
\displaystyle |M_n|^2 = M_0 + M_{s}+M_{s^{\prime}}+M_{s s^{\prime}}. \label{Eq3.1}
\end{eqnarray}
The calculation of the spin-independent terms $M_0$ is performed by using the standard technique (see e.g. \cite{Nikishov-1_JETP_64, Nikishov-3_JETP_64, BLP}) taking into account Eqs.(\ref{Eq2.65}), (\ref{Eq4}). It leads to the following expression:
\begin{eqnarray}
&\displaystyle M_{0} = 2 \pi e^2 m^2 \Big (-4 J_n^2 + \xi^2 (J_{n+1}^2 + J_{n-1}^2 -2 J_n^2) \cr & \displaystyle \times \Big (\frac{(p^{\prime}k)}{(pk)} + \frac{(pk)}{(p^{\prime}k)}\Big )\Big ).
 \label{Eq3.2}
\end{eqnarray}
One can see that multiplying this result by the factor 2 we obtain the ordinary formula for the process with unpolarized particles (see \cite{Nikishov-3_JETP_64, BLP}). 

The one-spin-dependent terms in the square of the matrix element are found as (some details of calculations are provided in Appendix A):
\begin{eqnarray}
& \displaystyle M_{s} = 2 \pi e^2 m^3 \frac{\xi^2 (k^{\prime}k)}{n(pk)(p^{\prime}k)} (J_{n-1}^2 - J_{n+1}^2) \cr & \displaystyle \times \Big ((sk^{\prime}) + (sk) \Big ( n - \frac{m^2 \xi^2 (k^{\prime}k)}{2(pk)^2} - \frac{m_\star^2 (k^{\prime}k)}{(pk) (p^{\prime}k)}\Big )\Big ), \cr
& \displaystyle M_{s^{\prime}} = 2 \pi e^2 m^3 \frac{\xi^2 (k^{\prime}k)}{n(pk)(p^{\prime}k)} (J_{n-1}^2 - J_{n+1}^2) \cr & \displaystyle \times \Big ((s^{\prime}k^{\prime}) + (s^{\prime}k)\Big (n - \frac{m^2 \xi^2 (k^{\prime}k)}{2(p^{\prime}k)^2} - \frac{m_\star^2 (k^{\prime}k)}{(pk) (p^{\prime}k)}\Big )\Big ) \label{Eq3.37}
\end{eqnarray}
It is the difference of these terms that determine the presence (or absence) of the self-polarization effect (see Sec.\ref{Sect4}).

Calculation of the term $M_{s s^{\prime}}$ depending on both spins yields:
\begin{eqnarray}
& \displaystyle M_{s s^{\prime}} = \frac{2 \pi e^2 m^2 \xi^2}{(pk) (p^{\prime}k)} (J_{n-1}^2 + J_{n+1}^2) \Big [-\frac{(k^{\prime}k)^2}{(pk) (p^{\prime}k)}\times \cr & \displaystyle \Big ( \frac{2 n (p^{\prime}k)(pk)}{(k^{\prime}k)} - m_{\star}^2\Big )(s^{\prime} k)(sk) + 2 (pk) (s^{\prime}k)(sp^{\prime}) - \cr & \displaystyle - 2 (pk)(p^{\prime}k) (s^{\prime}s) + 2(p^{\prime}k) (s^{\prime}p)(sk)\Big ] + \cr & \displaystyle + 8\pi e^2 J_n^2 \Big [m_{\star}^2 (s^{\prime}s) + \frac{(p^{\prime}k)}{(k^{\prime}k)}\frac{n (pk)(p^{\prime}k) - m_{\star}^2 (k^{\prime}k)}{2 n (pk)(p^{\prime}k) - m_{\star}^2 (k^{\prime}k)}\cr & \displaystyle \times \Big ((s^{\prime}k^{\prime})(sp^{\prime}) + (s^{\prime}p)(sk^{\prime})\Big ) - \frac{(p^{\prime}k)}{2(k^{\prime}k)}\Big (\frac{m^2 \xi^2 (k^{\prime}k)}{(pk)(p^{\prime}k)} + \cr & \displaystyle + n \frac{2n(pk)(p^{\prime}k) - m^2 \xi^2 (k^{\prime}k)}{2n(pk)(p^{\prime}k) - m_{\star}^2 (k^{\prime}k)}\Big ) (s^{\prime}k^{\prime})(sk) +\cr & \displaystyle + \Big (\frac{m^2 \xi^2 (k^{\prime}k)}{(pk) (p^{\prime}k)} + \frac{(p^{\prime}k)}{2(k^{\prime}k)}\Big (\frac{m^2 \xi^2 (k^{\prime}k)}{(pk)(p^{\prime}k)} + \cr & \displaystyle + n \frac{2n(pk)(p^{\prime}k) - m^2 \xi^2 (k^{\prime}k)}{2n(pk)(p^{\prime}k) - m_{\star}^2 (k^{\prime}k)}\Big ) \Big ) (s^{\prime}k)(sk^{\prime}) - \cr & \displaystyle - \frac{n m^2 (k^{\prime}k)}{2 n (p^{\prime}k)(pk) - m_{\star}^2 (k^{\prime}k)} (s^{\prime}k)(sp^{\prime})\Big ]. \label{Eq3.68}
\end{eqnarray}
It is clear that this expression (as well as (\ref{Eq3.37})) may be represented in another form by using the conservation laws (\ref{Eq7}).

\section{Description of the final electron's polarization}
\label{Sect3}

The expression for the squared amplitude provided allows one to describe the polarization states of the electron in invariant form. For instance, the final formula for effective cross section of Ref.\cite{Serbo_EPJ_04} contains the expansions of the spin 4-vectors with the use of the following set of orthogonal unit 4-vectors:
\begin{eqnarray}
& \displaystyle s = \sum \limits_{j = 1}^3 \zeta_j n_j, s^{\prime} = \sum \limits_{j = 1}^3 \zeta_j^{\prime} n_j^{\prime}, \zeta_j = - (s n_j), \zeta_j^{\prime} = - (s^{\prime} n_j^{\prime}), \cr & \displaystyle n_1 = e_1, \ n_2 = -e_2 - k \frac{1}{(pk)}\sqrt{\frac{2 n (p^{\prime}k)(pk)}{(k^{\prime}k)} - m_{\star}^2 },\cr & \displaystyle n_3 = \frac{1}{m}\Big (p - k \frac{m^2}{(pk)}\Big ),\ \cr & \displaystyle n_1^{\prime} = e_1,\ n_2^{\prime} = -e_2 - k \frac{1}{(p^{\prime}k)}\sqrt{\frac{2 n (p^{\prime}k)(pk)}{(k^{\prime}k)} - m_{\star}^2 },\cr & \displaystyle n_3^{\prime} = \frac{1}{m}\Big (p^{\prime} - k \frac{m^2}{(p^{\prime}k)}\Big ). \label{Eq4.1}
\end{eqnarray}

The meaning of this expansion becomes clear from the fact that invariants $\zeta_1, \zeta_1^{\prime}$ characterize the electron's polarization perpendicular to the scattering plane, and the others characterize polarization in that plane. In particular, the invariants
\begin{eqnarray}
& \displaystyle \zeta_3 = \frac{m}{(pk)}(sk) = -({\bm \zeta}, {\bf n}_k),\cr & \displaystyle \zeta_3^{\prime} = \frac{m}{(p^{\prime}k)}(s^{\prime}k) = -({\bm \zeta}^{\prime}, \tilde{{\bf n}}_k) \label{Eq4.1a}
\end{eqnarray}
represent the projections of the electron's spin on the direction of propagation of the initial photon. Here, ${\bf n}_k = {\bf k}/\omega$ and $\tilde{{\bf n}}_k = \tilde{{\bf k}}/\omega$ are the unit vectors of quantization in the corresponding rest frames. For the usually-studied geometry of a head-on collision of a photon with an ultrarelativistic electron, one has $\zeta_3 = 2 \lambda_e$ and $\zeta_3^{\prime} \approx  2 \lambda_e^{\prime}$, since the scattering angle of the final electron is small in the laboratory frame of reference (see in more detail \cite{Serbo_EPJ_04}). In other words, the Lorentz-invariant spin quantum number $\zeta_3^{\prime}$ coincides with the final electron's double mean helicity with an accuracy $\mathcal O(\gamma^{-2})$.

The choice of the spin invariants (\ref{Eq4.1a}) means that the spin quantization axis does not depend upon the final particles momenta. Such a description (which is similar to that of Ref.\cite{Ternov}) is equivalent to the one used in the theory of synchrotron radiation where the quantization vectors, ${\bf n}_H$ and $\tilde{{\bf n}}_H$, coincide in the quasiclassical limit (i.e. when neglecting the electron's scattering). However, it is the deviation of invariant $\zeta_3^{\prime}$ from the exact double mean helicity of the final electron (arising due to the non-zero scattering angle) that leads to a non-zero self-polarization of the moderately relativistic electron beam. In other words, the non-Lorentz-invariant final electron's helicity $2 \lambda_e^{\prime} = ({\bm \zeta}^{\prime}, {\bf p}^{\prime})/|{\bf p}^{\prime}|$ does depend upon the integration variables in Eq.(\ref{Eq3.69}) that leads to the non-zero radiative polarization, as we know from Refs.\cite{Bagrov_Preprint, B1, B2}. We will return to these questions in the next Section.

For comparison of the results obtained in the previous Section with the ones of Ref.\cite{Serbo_EPJ_04}, it is necessary to express the scalar products of the form $(sk), (s^{\prime}k^{\prime})$ in Eqs.(\ref{Eq3.37}), (\ref{Eq3.68}) in terms of the invariants $\zeta_j, \zeta^{\prime}_j$. This yields
\begin{eqnarray}
& \displaystyle  (s s^{\prime}) = -\zeta_1 \zeta_1^{\prime} - \zeta_2 \zeta_2^{\prime} + \zeta_2 \zeta_3^{\prime} \frac{(k^{\prime}k)}{m(pk)}\times \cr & \displaystyle \sqrt{\frac{2 n (p^{\prime}k)(pk)}{(k^{\prime}k)} - m_{\star}^2 } - \zeta_3 \zeta_2^{\prime} \frac{(k^{\prime}k)}{m(p^{\prime}k)}\times \cr & \displaystyle \sqrt{\frac{2 n (p^{\prime}k)(pk)}{(k^{\prime}k)} - m_{\star}^2 } + \zeta_3 \zeta_3^{\prime} \Big (1 - \frac{(pk)}{(p^{\prime}k)} - \frac{(p^{\prime}k)}{(pk)}\cr & \displaystyle + \frac{(k^{\prime}k)}{m^2}\Big (n - \frac{m^2 \xi^2 (k^{\prime}k)}{2 (pk) (p^{\prime}k)}\Big )\Big ),\cr & \displaystyle (s^{\prime}k^{\prime}) = -\zeta_2^{\prime} \frac{(k^{\prime}k)}{(p^{\prime}k)}\sqrt{\frac{2 n (p^{\prime}k)(pk)}{(k^{\prime}k)} - m_{\star}^2 }\cr & \displaystyle + \zeta_3^{\prime} \frac{1}{m} \Big ((pk) \Big (n - \frac{m^2 \xi^2 (k^{\prime}k)}{2 (pk) (p^{\prime}k)}\Big ) - m^2  \frac{(k^{\prime}k)}{(p^{\prime}k)}\Big ),  \cr & \displaystyle (s k^{\prime}) = -\zeta_2 \frac{(k^{\prime}k)}{(pk)}\sqrt{\frac{2 n (p^{\prime}k)(pk)}{(k^{\prime}k)} - m_{\star}^2 } + \cr & \displaystyle + \zeta_3 \frac{1}{m} \Big ((p^{\prime}k) \Big (n - \frac{m^2 \xi^2 (k^{\prime}k)}{2 (pk) (p^{\prime}k)}\Big ) - m^2  \frac{(k^{\prime }k)}{(pk)}\Big ),\cr & \displaystyle (s p^{\prime}) = \zeta_2 \frac{(k^{\prime}k)}{(pk)} \sqrt{\frac{2 n (p^{\prime}k)(pk)}{(k^{\prime}k)} - m_{\star}^2 }\cr & \displaystyle + \zeta_3 \frac{(k^{\prime}k)}{m} \Big (\frac{m^2}{(pk)} + n - \frac{m^2 \xi^2 (k^{\prime}k)}{2 (pk) (p^{\prime}k)}\Big ),\cr & \displaystyle (s^{\prime} p) = -\zeta_2^{\prime} \frac{(k^{\prime}k)}{(p^{\prime}k)}\sqrt{\frac{2 n (p^{\prime}k)(pk)}{(k^{\prime}k)} - m_{\star}^2 }\cr & \displaystyle + \zeta_3^{\prime} \frac{(k^{\prime}k)}{m} \Big (-\frac{m^2}{(p^{\prime}k)} + n - \frac{m^2 \xi^2 (k^{\prime}k)}{2 (pk) (p^{\prime}k)}\Big ). \label{Eq4.4}
\end{eqnarray}
Substituting these expressions into (\ref{Eq3.37}), (\ref{Eq3.68}), one can obtain for the squared matrix element $|\mathcal M_n|^2 $ the corresponding formulas of Ref.\cite{Serbo_EPJ_04} with an accuracy up to the factor 2 (due to the differences in notations: see Eqs.(46), (49)-(51) in the work cited). Formulas for the effective cross section (our (\ref{Eq3.71}) and (32) of \cite{Serbo_EPJ_04}) also differ by the factor 2 due to summation over polarization states of the final photon we performed.

Though the expansion (\ref{Eq4.1}) may be used for description of the final electron's polarization, it seems to be relevant to derive the more general invariant representation for the matrix element squared. Such a representation turns out to be more convenient for analysis of the moderately relativistic case and a non-head-on collision geometry (where the radiative polarization takes place only). For this purpose, we write down the squared amplitude in the following form (when transforming 4-vector $F$, the conservation laws (\ref{Eq4}), (\ref{Eq5}), (\ref{Eq7}) were used):
\begin{eqnarray}
& \displaystyle |\mathcal M_n|^2 = F_0 + (F s^{\prime}),\cr & \displaystyle F_0 = \mathcal M_0 + \mathcal M_s,\ F^{\mu} = f_1 p^{\mu} + f_2 k^{\mu} + f_3 s^{\mu}, \label{Eq4.44}
\end{eqnarray}
and the invariant factors are found as:
\begin{eqnarray}
& \displaystyle F_0 = - \frac{2 J_n^2}{\xi^2} + (J_{n-1}^2 + J_{n+1}^2 - 2 J_n^2) \Big (\frac{(pk)}{2 (p^{\prime}k)} + \frac{(p^{\prime}k)}{2 (pk)}\Big )\cr & \displaystyle + \frac{m (k^{\prime}k)}{2n (pk)(p^{\prime}k)} (J_{n-1}^2 - J_{n+1}^2)\cr & \displaystyle \times \Big ((sk^{\prime}) + (sk) \Big ( n - \frac{m^2 \xi^2 (k^{\prime}k)}{2(pk)^2} - \frac{m_\star^2 (k^{\prime}k)}{(pk) (p^{\prime}k)}\Big )\Big ), \cr & \displaystyle f_1 = \frac{m (k^{\prime}k)}{2n(pk)(p^{\prime}k)} (J_{n-1}^2 - J_{n+1}^2) + \frac{(sk)}{(pk)}\Big (J_{n-1}^2 + \cr & \displaystyle + J_{n+1}^2 - 2 J_n^2 - \frac{2 J_n^2}{\xi^2} \frac{n (pk) (p^{\prime}k)}{2n (pk) (p^{\prime}k) - m_{\star}^2 (k^{\prime}k)}\Big ),  \cr & \displaystyle f_2 = \frac{m (k^{\prime}k)}{2n(pk)(p^{\prime}k)} (J_{n-1}^2 - J_{n+1}^2) \Big (2n - \frac{m_{\star}^2 (k^{\prime}k)}{(pk) (p^{\prime}k)} -\cr & \displaystyle \frac{m^2 \xi^2 (k^{\prime}k)}{2 (p^{\prime}k)} \Big (\frac{1}{(pk)} + \frac{1}{(p^{\prime}k)}\Big )\Big ) + \frac{J_{n-1}^2 + J_{n+1}^2 - 2 J_n^2}{(p^{\prime}k)}\cr & \displaystyle \times \Big ((sp^{\prime}) - (sk) \frac{(k^{\prime}k)(2n (pk)(p^{\prime}k) - m_{\star}^2 (k^{\prime}k))}{2(p^{\prime}k)(pk)^2} \Big ) + \cr & \displaystyle \frac{2 J_n^2}{\xi^2}\Big ((sk) \frac{m^2 \xi^2 (k^{\prime}k)^2}{2 (pk)^2 (p^{\prime}k)^2} - (sp^{\prime}) \frac{n (pk)}{2n (pk)(p^{\prime}k) - m_{\star}^2 (k^{\prime}k)}\Big ), \cr & \displaystyle f_3 = \frac{2 J_n^2}{\xi^2} - (J_{n-1}^2 + J_{n+1}^2 - 2 J_n^2).\label{Eq4.45}
\end{eqnarray}
Note that from this formula one can easily derive the corresponding expression for the process of $e^{+}e^{-}$-pair creation by a photon propagating in the laser field (non-linear Breit-Wheeler process) by using the crossing symmetry; see e.g. \cite{Serbo_EPJ_05}. In the limiting case of the weak laser wave, this formula coincides with the one known in the literature for an ordinary Compton scattering (see Appendix B).

According to the usual rules \cite{BLP}, the 4-vector describing the electron spin state resulting from the scattering process itself is obtained from the representation (\ref{Eq4.44}) as follows
\begin{eqnarray}
&& \displaystyle s^{(f)}_\mu = \frac{F_{\mu}}{F_0}. \label{Eq4.46}
\end{eqnarray}
Accordingly, the 3-vector of the electron polarization resulting from the scattering process itself has the form:
\begin{eqnarray}
&& \displaystyle {\bm \zeta}^{(f)} = \frac{1}{F_0}\Big (f_1 {\tilde {\bf p}} + f_2 {\tilde {\bf k}} + f_3 {\tilde {\bf s}}\Big ). \label{Eq4.47}
\end{eqnarray}
Here, tilde denotes that the vector is taken in the rest frame of the final electron. Transforming all the vectors in Eq.(\ref{Eq4.47}) into laboratory frame of reference, we obtain the following expressions for the longitudinal polarization and transverse polarization of the final electron (we recall that prime stands for a particle in the final state):
\begin{eqnarray}
& \displaystyle \zeta^{(f)}_{\parallel } \equiv \Big (\bm \zeta^{(f)} \frac{{\bf p}^{\prime}}{|{\bf p}^{\prime}|}\Big ) = \frac{f_1}{F_0} \Big (\gamma^{\prime} \frac{ ({\bf p} {\bf p}^{\prime})}{|{\bf p}^{\prime}|} - \gamma |{\bf p}^{\prime}|\Big ) + \cr & \displaystyle + \frac{f_2}{F_0} \Big (\gamma^{\prime} \frac{ ({\bf k} {\bf p}^{\prime})}{|{\bf p}^{\prime}|} - \frac{\omega}{m} |{\bf p}^{\prime}|\Big ) +\cr & \displaystyle + \frac{f_3}{F_0} \Big (\zeta_{\parallel } \Big (\gamma \gamma^{\prime} \frac{({\bf p}^{\prime} {\bf p})}{|{\bf p}^{\prime}| |{\bf p}|} - \frac{|{\bf p}^{\prime}|}{m} \frac{|{\bf p}|}{m}\Big ) + \gamma^{\prime} \frac{(\bm \zeta_{\perp }{\bf p}^{\prime})}{|{\bf p}^{\prime}|}\Big ), \cr & \displaystyle
\bm \zeta_{\perp}^{(f)} \equiv  - \frac{{\bf p}^{\prime}}{|{\bf p}^{\prime}|} \times \Big [\frac{{\bf p}^{\prime}}{|{\bf p}^{\prime}|}\times \bm \zeta^{(f)} \Big ] =\cr & \displaystyle = - \frac{1}{F_0} \frac{{\bf p}^{\prime}}{|{\bf p}^{\prime}|} \times \Big [\frac{{\bf p}^{\prime}}{|{\bf p}^{\prime}|} \times \Big (f_1 {\bf p} + f_2 {\bf k} + \cr & \displaystyle + f_3 \Big (\frac{\bf p}{|{\bf p}|}\gamma \zeta_{\parallel} + \bm \zeta_{\perp } \Big ) \Big ) \Big ]. \label{Eq4.48}
\end{eqnarray}
Here, the space components of the initial electron polarization 4-vector ${\bf s}$ were expanded according to Eq.(\ref{Eq6}), and all the 3-vectors (except $\bm \zeta$) are defined in the laboratory frame of reference.

\section{Analysis of the process with a spin flip}
\label{Sect4}

As the first step, let us demonstrate that the radiative polarization effect is absent when the spin quantization axis is chosen along the initial photon momentum according to invariants (\ref{Eq4.1a}). In this case one may put $\zeta_{1, 2} = \zeta_{1, 2}^{\prime} = 0$, and the terms linear in the spin quantum numbers in the squared amplitude (\ref{Eq4.44}) have the following form:
\begin{eqnarray}
&\displaystyle |\mathcal M_n|^2 \propto \mathcal M_s + \mathcal M_{s^{\prime}} = a \zeta_3 + b \zeta_3^{\prime},\cr &\displaystyle a = b = \frac{(k^{\prime}k)}{2 n (pk) (p^{\prime}k)} (J_{n-1}^2 - J_{n+1}^2) \times \cr & \displaystyle \Big ((pk) + (p^{\prime}k)\Big ) \Big (n - \frac{m_{\star}^2 (k^{\prime}k)}{(pk)(p^{\prime}k)}\Big ). \label{Eq5.1}
\end{eqnarray}
Since the coefficients of the spin invariants coincide, the probabilities $W_{\downarrow \uparrow }$ and $W_{ \uparrow \downarrow}$ of the process with a spin flip ($\zeta_3^{\prime} = - \zeta_3$) are equal to each other, and so $P = 0$. This conclusion is in agreement with \cite{Ternov, Serbo_PR_03}. 

However, the Lorentz-invariant spin quantum number $\zeta_3^{\prime}$ coincides with the non-Lorentz-invariant double helicity only in the ultrarelativistic case (see in more detail \cite{Serbo_EPJ_04}). Hence, the more accurate consideration is required for a moderately relativistic electron and geometry of the non-head-on collision. In fact, the zero result (\ref{Eq5.1}) is correct only when neglecting the electron's scattering, as it is easy to see directly from Eq.(\ref{Eq3.37}) (see the corresponding footnote in Ref.\cite{Serbo_PR_03}). In this case the electron almost conserves its energy and momentum, so in the linear regime the spin-dependent functions, $\mathcal M_s$ and $ \mathcal M_{s^{\prime}}$, are equal to each other with an accuracy $\mathcal O(\theta_e^2)$ ($\theta_e$ is the electron's scattering angle)\,\footnote{As it follows from Eq.(\ref{Eq3.37}), these functions would be equal to each other if the spin scalar products (e.g. $(sk)$ and $(s^{\prime}k)$) would be equal too. For the spins being quantized along the initial photon's momentum in the laboratory frame (${\bm \zeta} = \zeta {\bf k}/\omega, {\bm \zeta}^{\prime} = \zeta^{\prime} {\bf k}/\omega, \zeta = \zeta^{\prime} = 1, {\bf k}/\omega = - {\bf p}/|{\bf p}|$), these products in the same frame are: $(sk) = -\omega |{\bf p}|/m + ({\bf k}, {\bf p})E/(|{\bf p}| m), (s^{\prime}k) = -\omega |{\bf p}^{\prime}|\cos \theta_e/m + ({\bf k}, {\bf p^{\prime}})E^{\prime}\cos \theta_e/(|{\bf p}^{\prime}| m) + ({\bf k}, {\bf p}/|{\bf p}| - {\bf p^{\prime}} \cos \theta_e/|{\bf p}^{\prime}|)$. Thus, these expressions are equal to each other only when neglecting the electron's scattering: $\theta_e \rightarrow 0, |{\bf p}^{\prime}| \approx |{\bf p}|$.}. 

For a head-on collision of the electron with $\gamma \sim 5$ and the photon with $\omega \sim 1$ eV, the difference between them seems to be negligibly small (for the almost back-scattered final photon and the spins being quantized along the initial photon's momentum in the laboratory frame of reference): 
\begin{eqnarray}
&&\displaystyle \mathcal M_s - \mathcal M_{s^{\prime}} \sim \theta_e^2 \lesssim \left (\frac{2 \omega}{m}\right )^2 \sim 10^{-11}. \label{Eq5.2aa}
\end{eqnarray}
However, in order to estimate the polarization degree on the amplitude level one should divide this quantity by a sum 
\begin{eqnarray}
&&\displaystyle \mathcal M_0 + \mathcal M_{s s^{\prime}(\uparrow \downarrow) } \lesssim 10^{-10} \label{Eq5.2ab}
\end{eqnarray}
which turns out to be small as well (for the same parameters). Accordingly, the ratio $(|\mathcal M_n|^2_{\downarrow \uparrow } - |\mathcal M_n|^2_{\uparrow \downarrow })/(|\mathcal M_n|^2_{\downarrow \uparrow } + |\mathcal M_n|^2_{\uparrow \downarrow })$ may achieve the values $\approx 0.1$ in the moderately relativistic regime. 

Let us illustrate this in more detail. For these purposes we will integrate the scattering amplitude squared (\ref{Eq4.44}) in the laboratory frame of reference in which the electron's spin quantum number represents helicity. In this frame the angle between the initial particles momenta is $\alpha$ ($\alpha = \pi$ corresponds to the head-on collision), and the z-axis coincides with direction of the initial photon momentum. The kinetic 3-momenta are:
\begin{eqnarray}
&\displaystyle {\bf k} = \omega \{ 0, 0, 1\},\ {\bf p} = m \sqrt {\gamma^2 - 1} \{ 0, \sin \alpha, \cos \alpha\},\cr & \displaystyle {\bf k}^{\prime} = \omega^{\prime}\{\sin \theta \sin \phi, \sin \theta \cos \phi, \cos \theta \}. \label{Eq5.2a}
\end{eqnarray}

In principle, it is possible to derive the exact analytical expression for the total probability with a spin flip (integrating, for instance, Eq.(\ref{Eq3.73a})), but the required calculations are rather cumbersome and the final result will not have the explicitly invariant form. This is due to the fact that we use non-Lorentz-invariant helicities as the spin quantum numbers (in contrast to Refs.\cite{Ternov, Serbo_PR_03}). That is why we will integrate the squared amplitude in Eq.(\ref{Eq3.73a}) numerically using the package Mathematica. Since the calculated values of the probabilities with spin-flips turn out to be rather small (see also \cite{E}), we use several methods of numerical integration whose predictions for polarization degree differ from each other within the accuracy 5\,\%.
Note that the compact analytical formulas for total probability were obtained in Refs.\cite{Bagrov_Preprint, B1, B2}. However, these formulas refer to the very special choice of the spin operators (see below).
\begin{figure}
\center
\includegraphics[width=6.50cm, height=5.00cm]{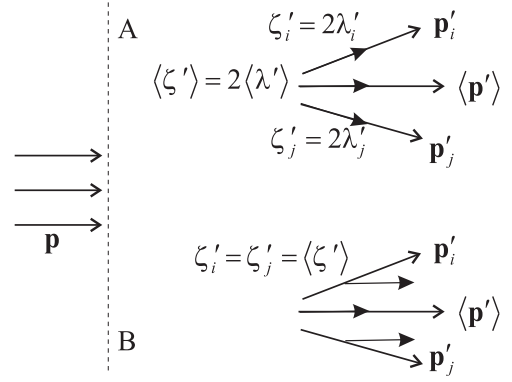}
\caption{\label{Fig1} Two different ways to describe polarization of the final electron (positron) beam having some momentum spread.}
\end{figure}

Describing the average longitudinal polarization of the final electron beam, it is necessary to distinguish two different ways. According to the first, the spin quantization axis is chosen along the final electron's momentum in the laboratory frame (description A; see Fig.1). So the final electron's spin quantum number is a mean helicity, and the spin 3-vectors in (\ref{Eq6}) have the form:
\begin{eqnarray}
&\displaystyle \bm \zeta = \frac{{\bf p}}{|{\bf p}|}2 \lambda_e,\ \bm \zeta^{\prime} = \frac{{\bf p}^{\prime}}{|{\bf p}^{\prime}|} 2 \lambda_e^{\prime}. \label{Eq5.2}
\end{eqnarray}
The spin flip occurs when $\lambda_e^{\prime} = - \lambda_e$. Note that in this case the final electron's spin quantization axis does depend upon the integration variables in Eq.(\ref{Eq3.73a}). 

However, in practice any beam has some angular divergence, so the alternative variant to choose the spin quantum number may turn out to be more convenient. Namely, the final electron's spin may be quantized along the direction of the average momentum of the beam in the laboratory frame (description B; see Fig.1). In this case the final electron's spin quantum number $\zeta^{\prime}$ is also a non-Lorentz-invariant quantity and the spin 3-vectors become
\begin{eqnarray}
&& \displaystyle \bm \zeta = \frac{{\bf p}}{|{\bf p}|} \zeta,\ \bm \zeta^{\prime} = \frac{{\bf p}}{|{\bf p}|}\zeta^{\prime},\ \zeta = 2 \lambda_e,\ \zeta^{\prime}\ne 2 \lambda_e^{\prime}.\label{Eq5.2b}
\end{eqnarray}
The spin flip occurs when $\zeta^{\prime} = - \zeta$. As the relativistic electron's scattering angle in this frame is small, both of these descriptions lead to the similar results, as we will show.

\begin{figure*}
\center
\includegraphics[width=15.00cm, height=5.00cm]{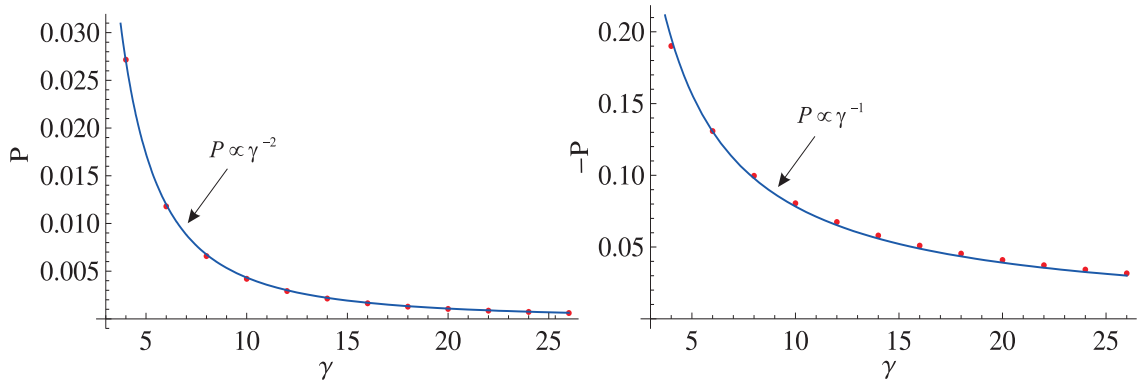}
\caption{\label{Fig2} Polarization of the final electron beam as a function of the initial electron's energy (weak laser wave, head-on collision, $n = 1, \alpha = \pi, \omega = 1$ eV). Left panel: description A. Right panel: description B. Dots: results of numerical calculations, curves: fits by the corresponding polynomials. }
\end{figure*}

\begin{figure}
\center
\includegraphics[width=7.00cm, height=4.50cm]{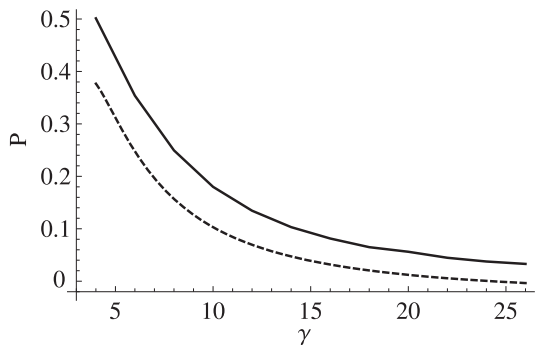}
\caption{\label{Fig3} The same as in Fig.\ref{Fig2}, but for a non-head-on collision ($n = 1, \alpha = \pi/6, \omega = 1$ eV). Solid curve: description A. Dashed curve: description B.}
\end{figure}

\begin{figure*}
\center
\includegraphics[width=16.00cm, height=5.50cm]{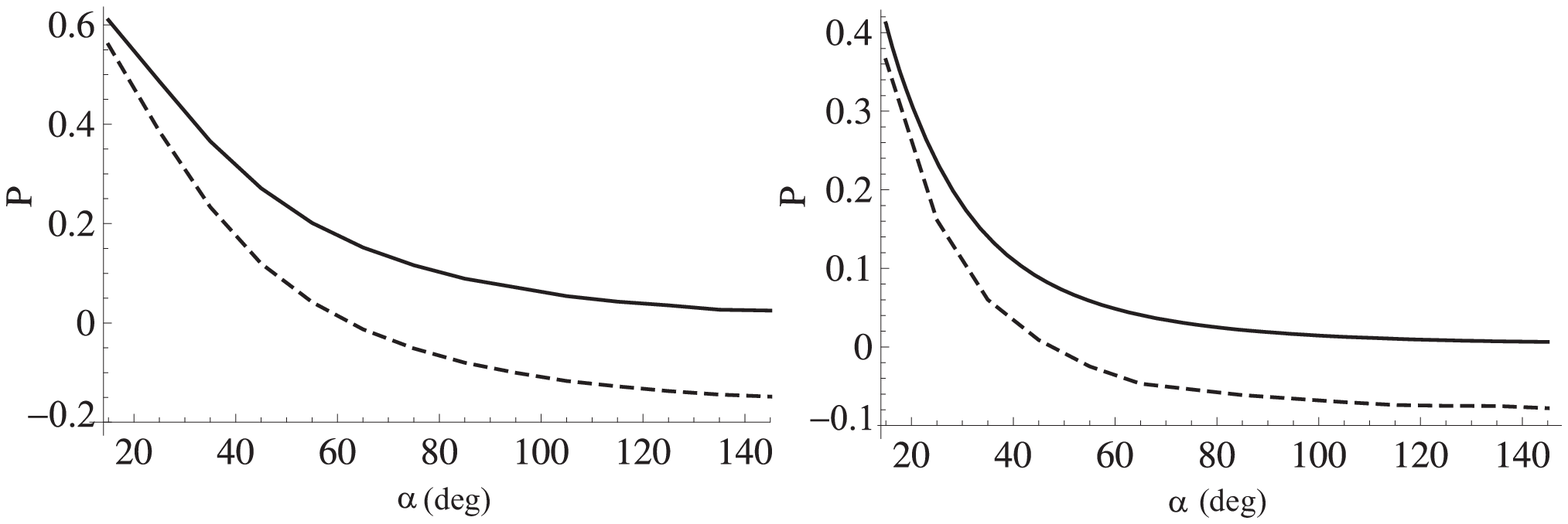}
\caption{\label{Fig4} Polarization of the final electron beam as a function of the collision angle (weak laser wave, $n = 1, \omega = 1$ eV). Left panel: $\gamma = 5$. Right panel: $\gamma = 10$. Solid curve: description A. Dashed curve: description B.}
\end{figure*}

First of all, one can easily verify that in the ultrarelativistic limit we recover the result $P \rightarrow 0$ according to both of the descriptions. Fig.2 shows the dependences of the degree of polarization (defined in (\ref{Eq5.8b})) for a head-on collision ($\alpha = \pi$) upon the energy of the initial electron in the case of a weak laser wave: $|{\bf E}|/E_c \sim 10^{-10}$ (linear Compton scattering). As it is clear from the fits, they fall as $P \propto \gamma^{-2}$ (description A) and as $P \propto \gamma^{-1}$ (description B). Therefore, the degree of polarization is small even for the moderately relativistic electrons. 

The situation changes for the better if the initial particles collide at the angle $\alpha \ll \pi$. Fig.3 shows the dependences similar to those of Fig.2, but for a non-head-on collision. While the degree of polarization stays small for $\gamma \gg 1$, it noticeably increases in the region of $\gamma \sim 5$ reaching $P \lesssim 0.5$ already at $\alpha = 30^{\circ}$. Fig.4 shows the dependences of $P$ upon the angle of collision for the different values of the initial electron's energy. As it follows from the plots, the degree of polarization as high as $P \approx 0.6$ (60\,\%) may be obtained by colliding the moderately relativistic electrons with $\gamma \lesssim 10$ and the photons of an optical or infrared laser at the angles $\alpha \ll \pi/2$. Note that for an electron with $\gamma \lesssim 10$ and a photon with $\omega \sim 1$ eV the invariant $x$ (defined in Eq.(\ref{Eq3.71})) stays rather small even for a head-on collision: $x \sim 10^{-5} - 10^{-4}$. It is this quantity that measures the electron's energy losses during the scattering process (see e.g. \cite{Serbo_PR_03}). Accordingly, for chosen calculation parameters the energy losses of the electron are less than a percent of its energy.

The further decrease of the collision angle or the electron's energy does not lead to the significant increase of the polarization degree. The maximum value of $P \approx 0.65$ (65\%) is observed for $\alpha \approx 20^{\circ}$ and $2 < \gamma < 5$ (in the description A). In addition, the probability of the process for very small collision angles ($\alpha \rightarrow 0$) is noticeably lower that results in the extremely high polarization times (see below). Note also that the polarization degree becomes even lower for non-relativistic electrons. 

In the non-linear regime with $\xi \lesssim 1$ \cite{SLAC} or even $\xi \gg 1$ (Vulcan laser facility \cite{Vulcan}, project ELI \cite{ELI}; see also a table in \cite{H}), the electron absorbs $n$ photons from the wave and emits only one final photon. Numerical estimations of the influence of the process non-linearity on the degree of polarization for parameters similar to those of the experiment at SLAC \cite{SLAC} ($|{\bf E}|/E_c \sim 10^{-6}$, $\xi \sim 0.5$ for a photon with $\omega \sim 1$ eV) show no noticeable deviations from the linear scattering case. The difference of $P$ from the values calculated in the linear regime is observed for the laser field strength $|{\bf E}|/E_c \gtrsim 10^{-5}$ and higher, but a significant increase of the polarization degree does not take place (see similar estimations for a high-power laser and ultrarelativistic electrons in Ref.\cite{E}). Moreover, in the super-strong laser fields the electron beam's energy losses significantly increase that makes such a polarization (even if it would take place in such a field) inconvenient from a practical point of view.

On the contrary, the probability of the process itself is significantly higher in the case of a high-power laser that results in the significant decrease of the polarization time $T_{pol}$ (defined in Eq.(\ref{Eq3.76})). Table 1 shows the characteristic relaxation times in the linear and non-linear regimes for the collision angle $\alpha = 20^{\circ }$. One may show that $T_{pol}$ as a function of $\alpha$ decreases when increasing the collision angle (since the probability of emission rises, see below), and it is almost one order lower already for the angle $\alpha = 30^{\circ }$. Nevertheless, it is clear that the times necessary for polarization of a beam with degree $P$ are many orders higher than the durations of the modern laser pulses (picoseconds) that makes the polarization of an electron (positron) beam with the use of this technique practically impossible. 

The polarization time as a function of the laser field strength and the photon frequency has the following form:
\begin{eqnarray}
&\displaystyle T_{pol} \propto T_0 \Big ( \frac{E_c}{|{\bf E}|}\Big )^2 \frac{m}{\omega}. \label{Eq5.5}
\end{eqnarray}

Note also that for almost co-propagating electrons and photons ($\alpha \rightarrow 0$), the value of $T_{pol}$ falls as the initial electron's energy decreases (in contrast to the dependence provided in Table 1). Such a behaviour of $T_{pol}$ easily follows from the general considerations (see e.g. $\S$101 in  \cite{BLP}). Indeed, in the case being considered two wave invariants $H^2 - E^2$ and $({\bf E}\cdot{\bf H})$ are equal to zero, and the probability of the process can be expressed through the dynamic invariant
\begin{eqnarray}
&\displaystyle \chi^2 = -\frac{e^2}{m^6} (F_{\mu \nu}p^{\nu})^2 = \left (\frac{|{\bf E}^{\prime}|}{E_c}\right )^2\label{Eq5.6}
\end{eqnarray}
only. Here, $F_{\mu \nu} = \partial_{\mu }A_{\nu} - \partial_{\nu }A_{\mu}$ is the strength tensor of the laser field, and prime stands for the rest frame of the initial electron.
This means that $|{\bf E}^{\prime}|$ must take place in the denominator of (\ref{Eq5.5}) rather than $|{\bf E}|$. For a head-on collision we have $|{\bf E}^{\prime}| \approx 2 \gamma |{\bf E}|$, so the polarization time has the damping factor $\gamma^2 \gg 1$ in the denominator. However, for small values of the collision angle the opposite dependence takes place: $|{\bf E}^{\prime}| \approx |{\bf E}|\sqrt{(1 - \beta)/(1 + \beta)}$. Thus, for almost co-propagating electrons and photons the polarization time increases drastically as the electron's energy rises. This explains the extremely low probability of the process and, accordingly, extremely high polarization time for small collision angles.

\begin{table}
\caption{Approximate polarization time for the linear scattering (left column: $\xi = 5 \times10^{-5}$) and for the non-linear scattering (right column: $\xi = 0.5$; summed over 7 harmonics) according to both of the descriptions (A, B). Parameters: $\alpha = 20^\circ, \omega = 1$ eV.} \label{table1}
\begin{center}
\begin{tabular}{|c|c|c|}
\hline
$\gamma$ & $T_{pol}$, s ($|{\bf E}|/E_c = 10^{-10}$) & $T_{pol}$, s ($|{\bf E}|/E_c = 10^{-6}$) \\
\hline
5 & $7 \times 10^{8}$ (A), $8 \times 10^{8}$ (B) & $6$ (A), $7$ (B)\\
\hline
10 & $5 \times 10^{8}$ (A), $6 \times 10^{8}$ (B) & $4$ (A), $4$ (B)\\
\hline
\end{tabular}
\end{center}
\end{table}

An additional remark concerning the strong laser fields with $|{\bf E}|/E_c > 10^{-5}$ is in order. First of all, for a process in such a strong field we must take into account additional diagrams, for example, the ones with emission of more than one photon (see e.g. \cite{Lot}). When propagating through the laser field emitted photons may create an $e^{-}e^{+}$-plasma that becomes crucial already at field intensities as low as $I \sim 10^{24}\ W{cm}^{-2}$ (that corresponds to $|{\bf E}|/E_c \sim 10^{-3}$) \cite{Ner}. Note that we actually mean the root-mean-square values of the field strength rather than the peak ones. Secondly, any real focused laser beam possesses longitudinal components of its electric and magnetic fields (even in the paraxial approximation, see e.g. \cite{N, F}). In contrast to the plane wave (\ref{Eq2}) being considered in this paper, such a field has a non-zero invariant $H^2 - E^2$ that results in the possibility of $e^{+}e^{-}$-pair creation from vacuum. The threshold of this process has been shown to depend strongly upon the laser parameters and has a value of $I \lesssim 10^{26}\ W cm^{-2}$ ($|{\bf E}|/E_c \sim 10^{-2}$)\cite{N-2}. Thus, for field strengths higher than $|{\bf E}| \sim 10^{-5} E_c$ the model of the given (quasi-classical) laser field becomes invalid.

Finally, let us discuss the possible alternative descriptions of the final electron's polarization. Along with two descriptions for longitudinal polarization (Eqs. (\ref{Eq5.2}), (\ref{Eq5.2b})), one can also measure the transverse polarization of the final electrons projecting the spins on the unit vectors ${\bf p}\times {\bf k}/(|{\bf p}|\omega)$ and ${\bf p^{\prime}}\times {\bf k}/(|{\bf p^{\prime}}|\omega)$ for initial and final electron, respectively. One may easily prove that the degree of such a polarization is exactly zero for arbitrary values of the electron's energy and collision angle (it follows already from the squared amplitude). 

It is also important to note that none of the descriptions being discussed provides the high values of the polarization degree in the ultrarelativistic case reported in Refs.\cite{Bagrov_Preprint, B1, B2}. This may be explained by the fact that the eigenvalue of the spin operator used in the papers cited represents neither helicity nor the transverse polarization. The very similar result of the high polarization degree in the ultrarelativistic case was obtained in the preprint \cite{Pap} by integrating the cross section in the rest frame of the initial electron. Moreover, one may show that the formula for polarization time calculated according to such an approach coincides with expression derived in \cite{Bagrov_Preprint, B1, B2} (we will not adduce these rather cumbersome calculations here). However, as it was demonstrated in Ref.\cite{Serbo_PR_03} (and it is confirmed by the present calculations) the choice of the spin quantum numbers of \cite{Pap} has no practical interest. Accordingly, the self-polarization effect predicted in Refs.\cite{Bagrov_Preprint, B1, B2} cannot be observed experimentally.

\section{Discussion and conclusion}
\label{Sect5}

In this paper, we have reanalyzed the problem of the radiative polarization of electrons in a strong laser wave. We derived an expression for the squared amplitude which coincides with the one of Ref.\cite{Serbo_EPJ_04} and is presented in the form which allows one to consider the case of the non-head-on collision and the moderately relativistic electron. By choosing the exact electron's helicity as the spin quantum number we have studied in detail the process with a spin flip and conclude that the radiative polarization of electrons (positrons) is possible in the moderately relativistic regime only: $\gamma \lesssim 10$. This conclusion generalizes the one of the Refs.\cite{Ternov, Serbo_PR_03} where the Lorentz-invariant spin quantum numbers were used that provided the zero polarization. 

However, the effective polarization time of the electron beam exceeds $1$\,s even for a high power optical or infrared laser with $\xi \sim 0.1$ ($|{\bf E}| \sim 10^{-6}\, E_c$). We suppose that the maximum field strength for which the model of the given quasi-classical laser field stays valid is only one order higher: $|{\bf E}| \sim 10^{-5}\, E_c$ (the corresponding relaxation time is two orders lower). This makes the experimental realization of such a polarization technique practically impossible.

We also explained that the results of \cite{Serbo_PR_03} and \cite{Bagrov_Preprint, B1, B2} do not agree because the different spin quantum numbers were used in these papers. In particular, it seems that the choice which is made in Refs.\cite{Bagrov_Preprint, B1, B2} is inconvenient from a practical point of view.

It is important to note that our calculations refer to the emission rate of only one electron that results in the absence of any collective effects. One may expect some decrease of the polarization time during the \textit{coherent regime} of emission which is not described by the present model (up to several orders of magnitude; see the corresponding estimations for coherent synchrotron radiation in \cite{Pap_SR}). Coherent radiation of an electron beam occurs when the emitted photon wavelength becomes larger than the effective beam length: $\lambda^{\prime} \gtrsim l_B$. This indicates the region where the one-electron model of this paper becomes invalid. On the other hand, even regime of \textit{partial} coherence ($\lambda^{\prime} \lesssim l_B$) would not change noticeably the main conclusions of this paper. For a typical bunch length of $l_B \sim 1$\,mm and laser photons with $\omega \sim 1$\,eV (note that $\omega^{\prime} \sim \omega$ for a moderately relativistic electron and a non-head-on collision) condition $\lambda^{\prime} \ll l_B$ is always fulfilled because for such a bunch length radiation becomes coherent in the THz part of the spectrum. Nevertheless, even in the coherent regime of emission the values of polarization time stay many orders higher than durations of the modern laser pulses (picoseconds).

In conclusion, we would like to note that the model of the laser field used in this paper represents the simplest transverse plane wave. It is intuitively clear that the more realistic description of a focused laser pulse would provide the higher influence of the laser wave on the electron's spin. For example, an additional influence may be due to the non-zero longitudinal component of magnetic field of the focused laser pulse (the so-called TE-wave; see an exact solution of Maxwell equations presented in \cite{N, F}). However, an exact solution of the Dirac equation for an electron in such a field is unknown, which hampers the direct application of the Furry picture concept to this process. For a weak tightly focused laser wave one may expect that the corresponding calculations can be performed by generalizing the wave-packets approach used for Bessel photon beams carrying orbital angular momentum \cite{Ivanov}.

\begin{acknowledgments}
Author is grateful to V.G. Bagrov, A.P. Potylitsyn, V.G. Serbo and A.A. Tishchenko for stimulating criticism and useful discussions. The work is partially supported by the Russian Ministry for Science and Education within the program ``Personnel of Innovative Russia'' (contract No.$\Pi 1199$) and the internal TPU grant No.$2.81.11$.
\end{acknowledgments}

\

\appendix

\section{Some details of the squared matrix element calculations.}

When calculating the one-spin-dependent terms $M_s, M_{s^{\prime}}$, it is necessary to evaluate the products $\varepsilon^{\mu \nu \eta \rho}s_{\mu}k_{\nu}{k_{\eta}}^{\prime}e_{1 \rho}$. The last are calculated with the use of Eq.(\ref{Eq2.5}) and the following formula \cite{L-2}:
\begin{eqnarray}
\displaystyle && 
\varepsilon^{\mu \nu \eta \rho} \varepsilon_{\rho \alpha \beta \gamma} =  \left | %
\begin {array}{ccc}
\delta_{\alpha}^{\mu}& \delta_{\beta}^{\mu}& \delta_{\gamma}^{\mu}\\ 
\delta_{\alpha}^{\nu}& \delta_{\beta}^{\nu}& \delta_{\gamma}^{\nu}\\
\delta_{\alpha}^{\eta}& \delta_{\beta}^{\eta}& \delta_{\gamma}^{\eta}\\
\end {array} %
\right |. \label{Eq3.25}
\end{eqnarray}

When calculating the terms $M_{s s^{\prime}}$ depending upon both spins, it is necessary to evaluate the traces of six Dirac matrices that may be reduced to the traces of four matrices by using the standard anti-commutativity relation:
\begin{eqnarray}
& \displaystyle \Tr \{\gamma^{\mu} \gamma^{\eta} \gamma^{\rho} \gamma^{\sigma} \gamma^{\lambda} \gamma^{\tau}\} = g^{\mu \eta} \Tr \{\gamma^{\rho} \gamma^{\sigma} \gamma^{\lambda} \gamma^{\tau}\} - \cr & \displaystyle - g^{\mu \rho} \Tr \{\gamma^{\eta} \gamma^{\sigma} \gamma^{\lambda} \gamma^{\tau}\} + g^{\mu \sigma} \Tr \{\gamma^{\eta} \gamma^{\rho} \gamma^{\lambda} \gamma^{\tau}\} - \cr & \displaystyle - g^{\mu \lambda} \Tr \{\gamma^{\eta} \gamma^{\rho} \gamma^{\sigma} \gamma^{\tau}\} + g^{\mu \tau} \Tr \{\gamma^{\eta} \gamma^{\rho} \gamma^{\sigma} \gamma^{\lambda}\}, \label{Eq3.12}
\end{eqnarray}
or for the scalar products $\hat{a} \equiv (\gamma a)$:
\begin{eqnarray}
& \displaystyle \Tr \{\hat a\hat b\hat c\hat d\hat e \hat f\} = (a b) \Tr \{\hat c \hat d \hat e \hat f\} - (a c) \Tr \{\hat b \hat d \hat e \hat f\} \cr & \displaystyle + (a d) \Tr \{\hat b \hat c \hat e \hat f\} - (a e) \Tr \{\hat b \hat c \hat d \hat f\} + (a f) \Tr \{\hat b \hat c \hat d \hat e\}. \label{Eq3.13}
\end{eqnarray}

As a result, the part of the squared amplitude depending upon both spins reads
\begin{eqnarray}
& \displaystyle M_{s s^{\prime}} = \frac{2 \pi e^2 m^2 \xi^2}{(pk) (p^{\prime}k)} (J_{n-1}^2 + J_{n+1}^2) \Bigg [-\Big (\frac{(k^{\prime}k)^2}{(pk) (p^{\prime}k)}\times \cr & \displaystyle \Big ( \frac{2 n (p^{\prime}k)(pk)}{(k^{\prime}k)} - m_{\star}^2\Big ) + (k^{\prime}k) \Big (\frac{(pk)}{(p^{\prime}k)} + \frac{(p^{\prime}k)}{(pk)}\Big )\times \cr & \displaystyle \Big (n - \frac{m^2 \xi^2 (k^{\prime}k)}{2 (pk) (p^{\prime}k)}\Big )\Big )(s^{\prime} k)(sk) + \Big ((p^{\prime}k) + \frac{(pk)^2}{(p^{\prime}k)}\Big ) (s^{\prime}k)(sp^{\prime}) \cr & \displaystyle - (k^{\prime}k)^2\Big ((s^{\prime} e_1)(s e_1) + (s^{\prime} e_2)(s e_2)\Big ) - (s^{\prime}k)(s e_2)  \frac{(k^{\prime}k)^2}{(p^{\prime}k)}\times \cr & \displaystyle \sqrt{\frac{2 n (p^{\prime}k)(pk)}{(k^{\prime}k)} - m_{\star}^2} - (s^{\prime}e_2)(s k) \frac{(k^{\prime}k)^2}{(pk)}\times \cr &\displaystyle \sqrt{\frac{2 n (p^{\prime}k)(pk)}{(k^{\prime}k)} - m_{\star}^2} + \Big ((pk) + \frac{(p^{\prime}k)^2}{(pk)}\Big ) (s^{\prime}p)(sk) \cr & \displaystyle - ((p^{\prime}k)^2 + (pk)^2) (s^{\prime}s)\Bigg ] + \cr & \displaystyle + 8 \pi e^2 J_n^2 \Bigg [(s^{\prime}s) \Big (m_{\star}^2 + \frac{m^2 \xi^2 (k^{\prime}k)^2}{2(p^{\prime}k)(pk)}\Big ) -\cr & \displaystyle - (k^{\prime}k) \Big (n - \frac{m^2 \xi^2 (k^{\prime}k)}{2 (pk) (p^{\prime}k)}\Big ) \Big ((s^{\prime} e_1)(s e_1) + (s^{\prime} e_2)(s e_2)\Big ) \cr & \displaystyle + (s^{\prime}p)(sp^{\prime}) +  \frac{(s^{\prime}p)(s e_2)}{\sqrt{\frac{2 n (p^{\prime}k)(pk)}{(k^{\prime}k)} - m_{\star}^2}}\cr & \displaystyle \times \Big (m_{\star}^2 - n (p^{\prime}k) - n \frac{(p^{\prime}k)(pk)}{(k^{\prime}k)}\Big ) + \cr & \displaystyle +  \frac{(s^{\prime}e_2)(s p^{\prime})}{\sqrt{\frac{2 n (p^{\prime}k)(pk)}{(k^{\prime}k)} - m_{\star}^2}} \Big (m_{\star}^2 + n (pk) - n \frac{(p^{\prime}k)(pk)}{(k^{\prime}k)}\Big ) \cr & \displaystyle + n(k^{\prime}k) (s^{\prime}e_1)(s e_1) - (s^{\prime}p)(sk) n \frac{(pk)}{(k^{\prime}k)} - (s^{\prime}e_2)(sk)\times \cr & \displaystyle \frac{n (pk)}{\sqrt{\frac{2 n (p^{\prime}k)(pk)}{(k^{\prime}k)} - m_{\star}^2}} \Big (n - \frac{m^2 \xi^2 (k^{\prime}k)}{2 (pk) (p^{\prime}k)}\Big ) - \cr & \displaystyle - (s^{\prime}k)(s e_2)  \frac{n (p^{\prime}k)}{\sqrt{\frac{2 n (p^{\prime}k)(pk)}{(k^{\prime}k)} - m_{\star}^2}} \Big (n - \frac{m^2 \xi^2 (k^{\prime}k)}{2 (pk) (p^{\prime}k)}\Big ) \cr & \displaystyle - (s^{\prime}k)(s p^{\prime}) n \frac{(p^{\prime}k)}{(k^{\prime}k)}\Bigg ]. \label{Eq3.67}
\end{eqnarray}
This expression still depends upon the unit vectors $e_1, e_2$ through the scalar products of the form $(s^{\prime} e_1)(s e_1)$. Calculation of the last is performed by using the following formula \cite{L-2}:
\begin{eqnarray}
\displaystyle && 
\varepsilon^{\mu \nu \eta \rho} \varepsilon_{\alpha \beta \gamma \delta} = - 
\left | %
\begin {array}{cccc}
\delta_{\alpha}^{\mu}& \delta_{\beta}^{\mu}& \delta_{\gamma}^{\mu}& \delta_{\delta}^{\mu}\\ 
\delta_{\alpha}^{\nu}& \delta_{\beta}^{\nu} & \delta_{\gamma}^{\nu} & \delta_{\delta}^{\nu}\\
\delta_{\alpha}^{\eta}& \delta_{\beta}^{\eta}& \delta_{\gamma}^{\eta}& \delta_{\delta}^{\eta}\\
\delta_{\alpha}^{\rho}& \delta_{\beta}^{\rho}& \delta_{\gamma}^{\rho}& \delta_{\delta}^{\rho}\\
\end {array} %
\right |.\label{Eq3.28}
\end{eqnarray}
For the ``extra'' products in Eq.(\ref{Eq3.67}) we obtain:
\begin{eqnarray}
& \displaystyle (s^{\prime} e_1)(s e_1) = -(s^{\prime}s) + \frac{1}{(k^{\prime}k)^2}\Big (\frac{2 n (p^{\prime}k)(pk)}{(k^{\prime}k)} - m_{\star}^2 \Big )^{-1} \times \cr & \displaystyle \Big [(s^{\prime}k^{\prime})(sk) \Big ((k^{\prime}k)\Big (\frac{2 n (p^{\prime}k)(pk)}{(k^{\prime}k)} - m_{\star}^2 \Big ) \cr & \displaystyle - (p^{\prime}k)^2\Big (n - \frac{m^2 \xi^2 (k^{\prime}k)}{2 (pk) (p^{\prime}k)}\Big ) \Big ) - (s^{\prime}k)(sk)(p^{\prime}k)(pk) \times \cr & \displaystyle \Big (n - \frac{m^2 \xi^2 (k^{\prime}k)}{2 (pk) (p^{\prime}k)}\Big )^2 - (s^{\prime}k^{\prime})(sk^{\prime})(p^{\prime}k)(pk) \cr & \displaystyle + (s^{\prime}k)(sk^{\prime}) \Big ((k^{\prime}k)\Big (\frac{2 n (p^{\prime}k)(pk)}{(k^{\prime}k)} - m_{\star}^2 \Big ) - \cr & \displaystyle - (pk)^2\Big (n - \frac{m^2 \xi^2 (k^{\prime}k)}{2 (pk) (p^{\prime}k)}\Big ) \Big ) \Big ], \cr & \displaystyle
(s^{\prime}e_2) = -\frac{1}{(k^{\prime}k) \sqrt{\frac{2 n (p^{\prime}k)(pk)}{(k^{\prime}k)} - m_{\star}^2 }} \times \cr & \displaystyle \Big ((s^{\prime}k^{\prime})(p^{\prime}k) + (s^{\prime}k)(pk) \Big (n - \frac{m^2 \xi^2 (k^{\prime}k)}{2 (pk) (p^{\prime}k)}\Big ) \Big ), \cr & \displaystyle
(s e_2) = -\frac{1}{(k^{\prime}k) \sqrt{\frac{2 n (p^{\prime}k)(pk)}{(k^{\prime}k)} - m_{\star}^2 }} \times \cr & \displaystyle \Big ((s k^{\prime})(p k) + (s k)(p^{\prime}k) \Big (n - \frac{m^2 \xi^2 (k^{\prime}k)}{2 (pk) (p^{\prime}k)}\Big ) \Big ),
\label{Eq3.29}
\end{eqnarray}
that leads, in particular, to the rather simple expression:
\begin{eqnarray}
& \displaystyle (s^{\prime} e_1)(s e_1) + (s^{\prime} e_2)(s e_2) = \cr & \displaystyle = -(s^{\prime}s) + \frac{(s^{\prime}k^{\prime})(sk) + (s^{\prime}k)(sk^{\prime})}{(k^{\prime}k)}.\label{Eq3.30}
\end{eqnarray}

By using the six vectors we deal with, $s, s^{\prime}, p, p^{\prime}, k, k^{\prime}$, it is possible to construct only ten pairwise scalar products of the form $(s^{\prime}k)(sp^{\prime})$ (including $(s^{\prime}s)$ and taking into account that $(sp) = (s^{\prime}p^{\prime}) = 0$). Therefore, the terms $M_{s s^{\prime}}$ will be expressed through these ten products. Taking into account the ``spin conservation laws'' (\ref{Eq7}), the number of such products may be reduced to six. After the removal of the vectors $e_1, e_2$ from Eq.(\ref{Eq3.67}), one arrives at the more compact formula (\ref{Eq3.68}).

\section{The squared amplitude for the linear scattering}

In the limiting case of the weak laser wave ($\xi \rightarrow 0, \ n=1$) the combinations of Bessel functions which enter (\ref{Eq4.45}) have the following form (it follows from the asymptotes of these functions in the limit of the small argument):
\begin{eqnarray}
& \displaystyle J_{0}^2 - J_{2}^2 \rightarrow 1, \ J_{0}^2 + J_{2}^2 - 2 J_1^2 \rightarrow 1,\cr & \displaystyle \frac{2 J_1^2}{\xi^2} \rightarrow \frac{m^2 (k^{\prime}k)}{(pk)(p^{\prime}k)} \Big ( 1 - \frac{m^2 (k^{\prime}k)}{2(pk)(p^{\prime}k)} \Big ). \label{Eq4.457}
\end{eqnarray}
The corresponding invariant coefficients are:
\begin{eqnarray}
& \displaystyle F_0 \rightarrow \frac{(pk)}{2 (p^{\prime}k)} + \frac{(p^{\prime}k)}{2 (pk)} - \frac{m^2 (k^{\prime}k)}{(pk)(p^{\prime}k)} \Big (1 - \frac{m^2 (k^{\prime}k)}{2(pk)(p^{\prime}k)} \Big ) \cr & \displaystyle +  \frac{m (k^{\prime}k)}{2(pk)(p^{\prime}k)}\Big ((sk^{\prime}) + (sk) \Big (1 - \frac{m^2 (k^{\prime}k)}{(pk) (p^{\prime}k)}\Big )\Big ),\cr & \displaystyle f_1 \rightarrow \frac{1}{(pk)} \Big (\frac{m (k^{\prime}k)}{2(p^{\prime}k)} + (sk) \Big ( 1 - \frac{m^2 (k^{\prime}k)}{2(pk)(p^{\prime}k)} \Big )\Big ), \cr & \displaystyle f_2 \rightarrow \frac{1}{(p^{\prime}k)} \Big ( 1 - \frac{m^2 (k^{\prime}k)}{2(pk)(p^{\prime}k)} \Big ) \times \cr & \displaystyle \Big (\frac{m (k^{\prime}k)}{(pk)} + (sp^{\prime}) - (sk)\frac{(k^{\prime}k)}{(pk)}\Big ), \cr & \displaystyle f_3 \rightarrow \frac{m^2 (k^{\prime}k)}{(pk)(p^{\prime}k)} \Big ( 1 - \frac{m^2 (k^{\prime}k)}{2(pk)(p^{\prime}k)} \Big ) - 1.\label{Eq4.458}
\end{eqnarray}
Going to the rest frame of the initial electron ($p = \{m, \bold 0\}, s = \{0, \bm \zeta\}, \bold p^{\prime} = \bold k - \bold k^{\prime}$), we arrive at the following expressions entering the scattering amplitude squared:
\begin{eqnarray}
& \displaystyle 2 F_0 \rightarrow \frac{\omega}{\omega^{\prime}} + \frac{\omega^{\prime}}{\omega} - \sin^2 \theta - \frac{1-\cos \theta}{m} (\bm \zeta, \bold k \cos \theta + \bold k^{\prime}),\cr & \displaystyle 2 (Fs^{\prime}) \rightarrow - \frac{1-\cos \theta}{m} \Big (\bm \zeta^{\prime}, \bold k \cos \theta + \bold k^{\prime} - \cr & \displaystyle - (1 + \cos \theta ) (\bold k - \bold k^{\prime}) \frac{\omega + \omega^{\prime}}{\omega - \omega^{\prime} + 2m}\Big ) - \cr & \displaystyle - \frac{1 + \cos \theta}{m \omega^{\prime}} \Big ((\bm \zeta, \bold k - \bold k^{\prime}) (\bm \zeta^{\prime}, \bold k - \bold k^{\prime}) \times \cr & \displaystyle \frac{\omega (1 + \cos \theta ) + \omega^{\prime} \cos \theta (1 - \cos \theta )}{(1 + \cos \theta ) (\omega - \omega^{\prime} + 2m)} + \cr & \displaystyle + (\bm \zeta, \bold k) (\bm \zeta^{\prime}, \bold k^{\prime}) \frac{\omega (\omega - \omega^{\prime}) - 2 m \omega^{\prime}}{\omega (\omega - \omega^{\prime} + 2m)} + (\bm \zeta, \bold k^{\prime}) (\bm \zeta^{\prime}, \bold k) - \cr & \displaystyle - (\bm \zeta, \bold k) (\bm \zeta^{\prime}, \bold k) \frac{\omega^2 - (\omega^{\prime})^2}{\omega (\omega - \omega^{\prime} + 2m)}\Big ) + \cr & \displaystyle + (1 + \cos^2 \theta ) (\bm \zeta, \bm \zeta^{\prime}). \label{Eq4.459}
\end{eqnarray}
Here, the spin-independent terms in $F_0$ correspond to the Klein-Nishina formula, the one-spin-dependent terms coincide with expressions known in the literature (see, for example, functions $\bold f, \bold g$ in Eq.(87.23) of \cite{BLP} and also corresponding functions $\bold \Phi_i$ of \cite{L-T, T}). The terms depending on both spins in $(Fs^{\prime})$ also coincide with corresponding functions $\bold \Phi_2 (\bm \zeta, \bm \zeta^{\prime})$ of Refs.\cite{L-T, T} (see Eq.(2.11) and Eq.(4.16), respectively). Indeed, in the works cited one may find a sum $1 + \cos^2 \theta + \sin^2 \theta (\omega - \omega^{\prime})/(2m)$ at the product $(\bm \zeta, \bm \zeta^{\prime})$ where the last part of the sum seems to be superfluous. However, there also exists a term $-(\omega - \omega^{\prime})(\bm \zeta, \bold n \times \bold n^{\prime}) (\bm \zeta^{\prime}, \bold n \times \bold n^{\prime})/(2m) $ (where $\bold n = \bold k/\omega, \bold n^{\prime} = \bold k^{\prime}/\omega^{\prime}$) in the third line of Eq.(4.16) in \cite{T}. Evaluating then the necessary scalar products as 
\begin{eqnarray}
& \displaystyle
(\bm \zeta, \bold n \times \bold n^{\prime}) (\bm \zeta^{\prime}, \bold n \times \bold n^{\prime}) = \varepsilon_{ijk} \varepsilon_{mnl} \zeta_i \zeta^{\prime}_m n_j n_n n^{\prime}_k n^{\prime}_l 
\label{Eq4.45a}
\end{eqnarray}
and taking into account the following formula\cite{L-2}
\begin{eqnarray}
&& \displaystyle \varepsilon_{ijk} \varepsilon_{mnl} = \left | %
\begin {array}{ccc} 
\delta_{im}& \delta_{in}& \delta_{il}\\ 
\delta_{jm}& \delta_{jn}& \delta_{jl}\\
\delta_{km}& \delta_{kn}& \delta_{kl}\\
\end {array} %
\right |, \label{Eq4.4510}
\end{eqnarray}
we obtain, in addition, another term $(\bm \zeta, \bm \zeta^{\prime}) \sin^2 \theta$ \footnote{Author is grateful to V.\,G. Serbo for pointing out this circumstance.} which cancels the superfluous one in the first line of Eq.(4.16) in \cite{T}. Note that the same result for the square of the scattering amplitude in the linear regime is also given in \cite{Olsen}.


\begin{thebibliography}{50}

\bibitem{Nikishov-1_JETP_64}
A.I. Nikishov, V.I. Ritus, Sov. Phys. JETP \textbf {19}, 529 (1964).

\bibitem{G}
I.I. Goldman, Sov. Phys. JETP \textbf {19}, 954 (1964).

\bibitem{Nikishov-3_JETP_64}
N.B. Narozhny, A.I. Nikishov, V.I. Ritus, Sov. Phys. JETP \textbf {20}, 622 (1965).

\bibitem{BLP}
V.B. Berestetskii, E.M. Lifshitz, L.P. Pitaevskii, \textit{Quantum electrodynamics} (Oxford, Pergamon, 1982).

\bibitem{L-T}
F.W. Lipps, H.A. Tolhoek, Physica \textbf {20}, 395 (1954).

\bibitem{T}
H.A. Tolhoek, Rev. Mod. Phys. \textbf {28}, 277 (1956).

\bibitem{Serbo_NIMA_98}
G.L. Kotkin, S.I. Polityko, V.G. Serbo, Nucl. Instr. Meth. A \textbf {405}, 30 (1998).

\bibitem{Ternov}
I.M. Ternov, V.G. Bagrov, A.M. Khapaev, Ann. Phys. (Leipzig) \textbf {22}, 25 (1968).

\bibitem{G-2}
I.I. Goldman, V.A. Khoze, Phys. Lett. B {\bf 29}, 426 (1969); Sov. Phys. JETP {\bf 30}, 501 (1970).

\bibitem{Gr}
Ya.T. Grinchishin, M.P. Rekalo, Sov. Phys. JETP \textbf{57}, 935 (1983).

\bibitem{Pol}
E. Bol'shedvorsky, S. Polityko, A. Misaki, Progr. Theor. Phys. \textbf{104}, 769 (2000).

\bibitem{Serbo_EPJ_04}
D.Yu. Ivanov, G.L. Kotkin, V.G. Serbo, Eur. Phys. J. C \textbf{36}, 127 (2004).

\bibitem{Piazza}
A. Di Piazza, A.I. Milstein, C. M$\ddot{u}$ller, Phys. Rev. A \textbf{82}, 062110 (2010).

\bibitem{Muller}
T.-O. M$\ddot{u}$ller, C. M$\ddot{u}$ller, Phys. Lett. B \textbf{696}, 201 (2011).

\bibitem{Artru}
X. Artru, R. Chehab, M.Chevallier, et al., Nucl. Instr. and Meth. B \textbf{266}, 3868 (2008).

\bibitem{Zimm}
R.W. Assmann, F. Zimmermann, CERN SL-2001-064-AP; http://clic-study.web.cern.ch/CLIC-study/Publications/2001.html

\bibitem{B}
I.R. Bailey, Cockcroft-08-05; www.cockcroft.ac.uk/research/papers2008.htm

\bibitem{Klimenko}
Yu.I. Klimenko, O.S. Pavlova, B.A. Lysov, et al., Russ. Phys. J. \textbf{25}, 960 (1982).

\bibitem{Ternov-90}
I.M. Ternov, A.I. Studenikin, A.M. Khapaev, Russ. Phys. J. \textbf{33}, 45 (1990).

\bibitem{Bagrov_Preprint}
V.G. Bagrov, G.F. Kopytov, S.S. Oksuzyan, et al., Dep. VINITI, No. 4704 - B 88 (1988), in Russian.

\bibitem{B1}
V.G. Bagrov, N.I. Fedosov, G.F. Kopytov, et al., Il Nuovo Cimento B \textbf{103}, 549 (1989).

\bibitem{B2}
V.G. Bagrov, G.F. Kopytov, N.I. Fedosov, Yad. Fiz. \textbf{51}, 1336 (1990), in Russian.

\bibitem{Pap}
A.P. Potylitsyn, arXiv: 0203059v1.

\bibitem{Serbo_PR_03}
G.L. Kotkin, V.G. Serbo, V.I. Telnov, Phys. Rev. Spec. Top. - Accel. Beams \textbf{6}, 011001 (2003).

\bibitem{SLAC}
C. Bamber, S.J. Boege, T. Koffas, et al., Phys. Rev. D \textbf{60}, 092004 (1999).

\bibitem{Heinzl_10}
T. Heinzl, D. Seipt, B. K$\ddot{a}$mpfer, Phys. Rev. A \textbf{81}, 022125 (2010).

\bibitem{Serbo_EPJ_05}
D.Yu. Ivanov, G.L. Kotkin, V.G. Serbo, Eur. Phys. J. C \textbf{40}, 27 (2005).

\bibitem{E}
F. Ehlotzky, K. Krajewska, J.Z. Kami$\acute{n}$ski, Rep. Prog. Phys. \textbf{72}, 046401 (2009).

\bibitem{Vulcan} The Vulcan laser: www.clf.rl.ac.uk

\bibitem{ELI} The Extreme Light Infrastructure (ELI) project: www.eli-laser.eu

\bibitem{H}
T. Heinzl, J. Phys.: Conf. Ser. \textbf{198}, 012005 (2009).

\bibitem{Lot}
E. L$\ddot{o}$tstedt, U.D. Jentschura, Phys. Rev. Lett. \textbf{103}, 110404 (2009).

\bibitem{Ner}
E.N. Nerush, I.Yu Kostyukov, A.M. Fedotov, et al., Phys. Rev. Lett. \textbf{106}, 035001 (2011).

\bibitem{N}
N.B. Narozhny, M.S. Fofanov, JETP \textbf{90}, 753 (2000).

\bibitem{F}
A.M. Fedotov, K.Yu. Korolev, M.V. Legkov, Proc. SPIE \textbf{6726}, 672613 (2007).

\bibitem{N-2}
S.S. Bulanov, N.B. Narozhny, V.D. Mur, et al., JETP \textbf{102}, 9 (2006).

\bibitem{Pap_SR}
A.P. Potylitsyn, J. Phys. G: Nucl. Part. Phys. \textbf{37}, 115106 (2010).

\bibitem{Ivanov}
I.P. Ivanov, V.G. Serbo, Phys. Rev. A \textbf{84}, 033804 (2011).

\bibitem{L-2}
L.D. Landau, E.M. Lifshitz, \textit{The classical theory of fields} (Oxford, Pergamon, 1975).

\bibitem{Olsen}
H.A. Olsen, Applications of quantum electrodynamics, Springer tracts in modern physics \textbf{44}, 83 (1968).


\end{thebibliography}
\end{document}